\documentclass[12pt]{article}
\usepackage{epsfig}
\begin{document}
\baselineskip=15pt
\begin{titlepage}
\setcounter{page}{0}

\begin{center}
\vspace*{5mm}
{\Large \bf Observational Constraints on Variable Chaplygin Gas}\\
\vspace{15mm}

{\large Zong-Kuan Guo$^{b}$
 \footnote{e-mail address: guozk@itp.ac.cn}
 and Yuan-Zhong Zhang$^{a,b}$}\\
\vspace{10mm}
 {\it
 $^a$CCAST (World Lab.), P.O. Box 8730, Beijing 100080, China \\
 $^b$Institute of Theoretical Physics, Chinese Academy of
   Sciences, P.O. Box 2735, Beijing 100080, China}
\end{center}

\vspace{20mm}

\centerline{\large \bf Abstract}
 {We investigate observational constraints on the variable Chaplygin
gas model from the gold sample of type Ia supernova data and the
recent measurements of the X-ray gas mass fractions in galaxy
clusters. Combining these databases, we obtain a tight constraint
on the two model parameters. Our results indicate that the
original Chaplygin gas model is ruled out by the data at 99.7\%
confidence level.}

\vspace{2mm}

%\begin{flushleft}
%PACS number(s): 98.80.Es, 98.80.Cq
%\end{flushleft}

\end{titlepage}

%%========================section 1 =========================
\section{Introduction}

Recent observations of type Ia supernovae (SNe Ia) suggest that
the expansion of the universe is accelerating and that two-thirds
of the total energy density exists in a dark energy component with
negative pressure~\cite{AGR}. In addition, measurements of the
cosmic microwave background (CMB)~\cite{DNS} and the galaxy power
spectrum~\cite{MT} also indicate the existence of the dark energy.
The simplest candidate for the dark energy is a cosmological
constant $\Lambda$, which has pressure $P_\Lambda=-\rho_\Lambda$.
Specifically, a reliable model should explain why the present
amount of the dark energy is so small compared with the
fundamental scale (fine-tuning problem) and why it is comparable
with the critical density today (coincidence problem). The
cosmological constant suffers from these problems. One possible
approach to construct a viable model for dark energy is to
associate it with a slowly evolving and spatially homogeneous
scalar field $\phi$, called ``quintessence''~\cite{RP,ZWS}. Such a
model for a broad class of potentials can give the energy density
converging to its present value for a wide set of initial
conditions in the past and possess tracker behavior.

Recently, the Chaplygin gas model was proposed as an alternative
to the cosmological constant in explaining the accelerating
universe~\cite{KMP}. The Chaplygin gas is characterized by an
exotic equation of state $P=-A/\rho$, where $A$ is a positive
constant. An attractive feature of the model is that it can
naturally explain both dark energy and dark matter. The reason is
that the Chaplygin gas behaves as dust-like matter at early stage
and as a cosmological constant at later stage. Some possible
motivations for this model from the field theory point of view are
discussed in Refs.~\cite{BBS1}. The Chaplygin gas appears as an
effective fluid associated with $d$-branes~\cite{BH} and can also
be derived from the Born-Infeld action~\cite{BBS2}. An interesting
range of models was found to be consistent with SN Ia
data~\cite{MOW}, CMB experiments~\cite{BBS3} and other
observational data~\cite{DAJ}. The Chaplygin gas model has been
extensively studied in the literature~\cite{PFG}.

However, the Chaplygin gas model produces oscillations or
exponential blowup of the matter power spectrum that are
inconsistent with observation~\cite{STZW}. In Ref.~\cite{GZ}, we
considered a variable Chaplygin gas (VCG) model and showed that it
interpolates between a universe dominated by dust and a
quiessence-dominated one described by the constant equation of
state. Furthermore, we showed that the model corresponds to a
Born-Infeld tachyon action~\cite{GZ}. Recently, the model
parameters were constrained using the location of peaks of the CMB
spectrum and SN Ia data~\cite{SSKJD}. In this paper we consider
observational constraints on the VCG model from the gold sample of
157 SN Ia data and the recent measurements of the X-ray gas mass
fractions in 26 galaxy clusters. We perform a combined analysis of
these databases and obtain the confidence region on the two
parameters. Our results indicate that the original Chaplygin gas
is ruled out at $3\sigma$ confidence level.

%%=======================section 2===========================
\section{Variable Chaplygin Gas}

Let us now consider the VCG characterized by the equation of state
\begin{equation}
\label{es}
P_v=-\frac{A(a)}{\rho_v},
\end{equation}
where $A(a)$ is a positive function of the cosmological scale
factor $a$. This assumption is reasonable since $A(a)$ is related
to the scalar potential if we take the Chaplygin gas as a
Born-Infeld tachyon field~\cite{BBS2, GZ}. In a spatially flat
Friedmann-Robertson-Walker (FRW) universe, the energy conservation
equation is
\begin{equation}
\frac{d\rho_v}{da}=-3\,\frac{\rho_v +P_v}{a}.
\end{equation}
By inserting Eq.~(\ref{es}) into the above equation, one finds
that the VCG density evolves as
\begin{equation}
\label{ide}
\rho_v(a)=a^{-3}\left[6\int A(a)a^5da+B\right]^{1/2},
\end{equation}
where $B$ is an integration constant. Given a function $A(a)$,
Eq.~(\ref{ide}) allows us to obtain a solution $\rho_v(a)$. We
assume $A(a)$ is of the form $A(a)=A_0 a^{-n}$, where $A_0$ and
$n$ are constants. This ansatz has the following important
features: a) the VCG model with $n=0$ reduces to the original
scenario, b) we will see that the VCG behaves as a quiessence
rather than a cosmological constant at late times and c) the
function $\rho(a)$ can be calculated analytically. Then from
Eq.~(\ref{ide}) it follows that
\begin{equation}
\label{de}
\rho_v(a)=\sqrt{\frac{6}{6-n}\frac{A_0}{a^n}
 +\frac{B}{a^6}}\,.
\end{equation}
Note that $n=0$ corresponds to the original Chaplygin gas
scenario, in which the Chaplygin gas behaves initially as
dust-like matter and later as a cosmological constant~\cite{KMP}.
However, Eq.~(\ref{de}) shows that, in the VCG scenario, it
interpolates between a dust-dominated phase and a
quiessence-dominated phase described by the constant equation of
state $w=-1+n/6$~\cite{GOZ}. From Eq.~(\ref{de}) we get the
present value of the energy density of the VCG
\begin{equation}
\rho_{v0}=\sqrt{\frac{6}{6-n}A_0+B}\,,
\end{equation}
where the present value of the scale factor is normalized to
unity, i.e., $a_0=1$. Defining $B_s \equiv B/\rho_{v0}^2$,
Eq.~(\ref{de}) takes the form
\begin{equation}
\label{rhoz}
\rho_v(z)=\rho_{v0}\left[B_s(1+z)^6
 +(1-B_s)(1+z)^n\right]^{1/2},
\end{equation}
where $z=1/a-1$ is redshift. In the spatially flat FRW metric the
Friedmann equation can be written as
\begin{equation}
\label{fe}
H^2 = \frac{\kappa^2}{3}(\rho_b+\rho_v),
\end{equation}
where $H\equiv \dot{a}/a$ is the Hubble parameter, $\kappa^2\equiv
8\pi G$ is the gravitational coupling and $\rho_b$ is the energy
density of the baronic matter.  Substituting Eq.~(\ref{rhoz}) into
the Friedmann equation (\ref{fe}) gives
\begin{eqnarray}
H^2(z)/H_0^2 &=& \Omega_b(1+z)^3+(1-\Omega_b)\nonumber\\
 &&\times\left[B_s(1+z)^6+(1-B_s)(1+z)^n\right]^{1/2}\nonumber\\
 &\equiv& E^2(z;\Omega_b,B_s,n),
\end{eqnarray}
where $H_0\equiv 100h\,\mathrm{kms}^{-1}\mathrm{Mpc}^{-1}$ is the
present value of the Hubble parameter and $\Omega_b$ is the
density parameter of the baryonic matter component. Then it is
straightforward to show that the luminosity distance $d_L$ and the
angular diameter distance $d_A$ in the spatially flat FRW universe
are respectively given by
\begin{eqnarray}
d_L &=& \frac{c}{H_0}(1+z)\int_0^z\frac{dz}{E(z)}\,, \\
d_A &=& (1+z)^{-2} d_L.
\end{eqnarray}

%%=======================section 3===========================
\section{Supernova Ia Constraints}

\begin{figure}
\begin{center}
\includegraphics[width=10cm]{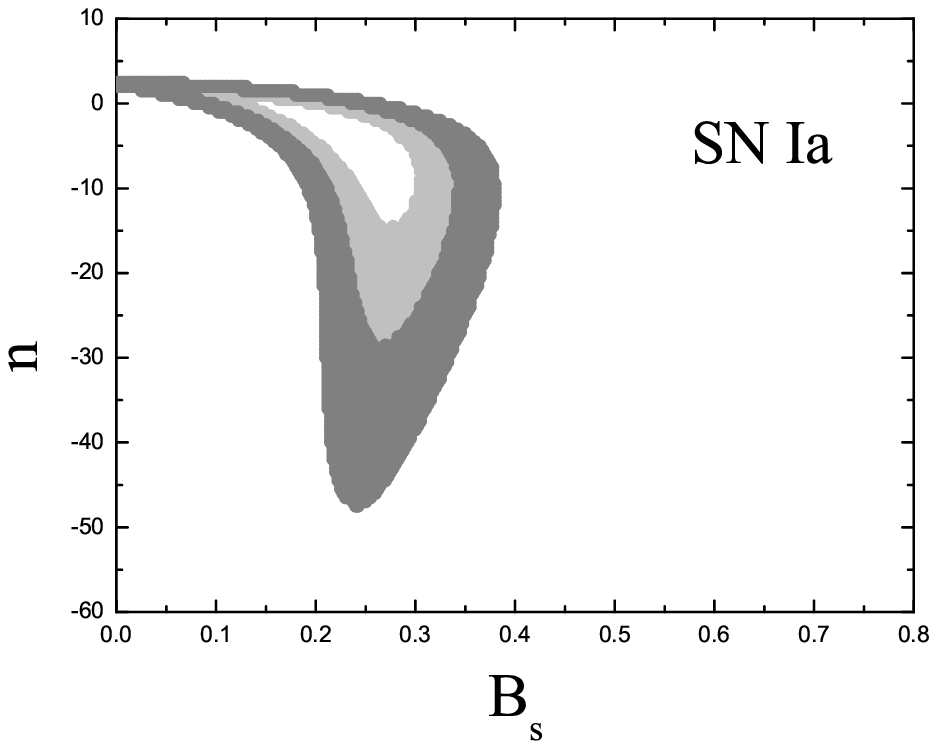}
\caption{Probability contours for $B_s$ versus $n$ are shown at
$1\sigma$, $2\sigma$, and $3\sigma$ when $\Omega_{\rm tot}=1$.
These constraints use the gold sample of 157 SN Ia
data~\cite{AR}.}
 \label{fig:SN}
\end{center}
\begin{center}
\includegraphics[width=10cm]{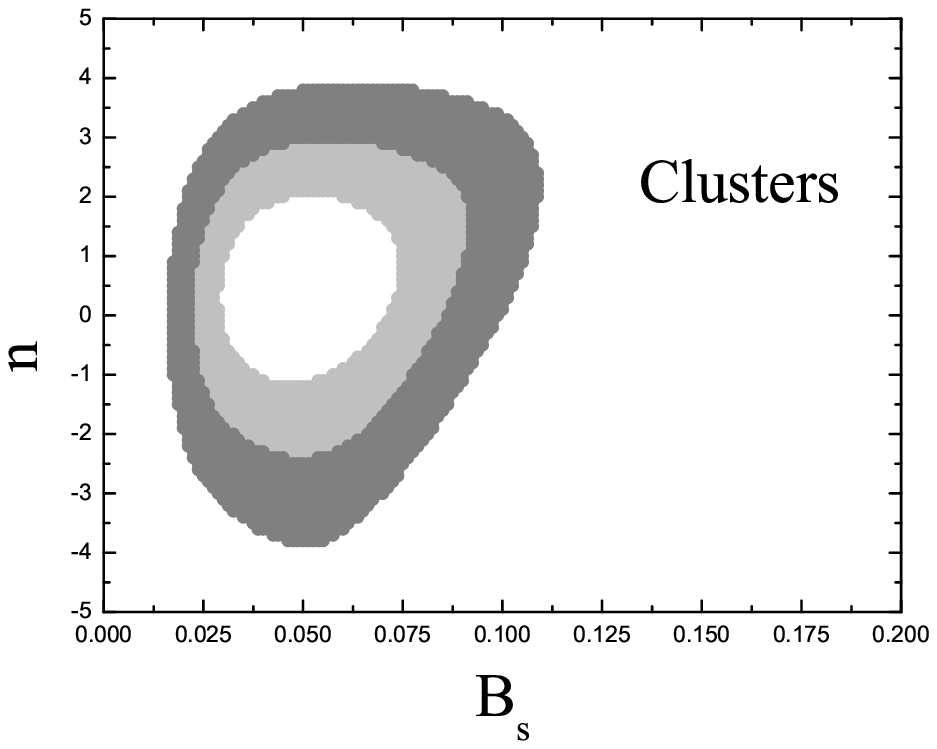}
\caption{Probability contours for $B_s$ versus $n$ are shown at
$1\sigma$, $2\sigma$, and $3\sigma$ when $\Omega_{\rm tot}=1$.
These constraints use the X-ray gas mass fraction in 26 galaxy
clusters ~\cite{ASEFL}.} \label{fig:Xray}
\end{center}
\end{figure}

Let us now consider constraints on the VCG model from the gold
sample of 157 SN Ia data compiled in Ref.~\cite{AR}. The
parameters in the model are determined by minimizing
\begin{equation}
\chi^2_{\mathrm{SN}}=\sum_{i=1}^{157}\frac{[\mu_\mathrm{mod}
(z_i;h,\Omega_b,B_s,n)-\mu_\mathrm{obs}(z_i)]^2} {\sigma_i^2},
\end{equation}
where $\sigma_i$ is the total uncertainty in the observation,
$\mu_\mathrm{obs}$ is the observed distance modulus of SNe Ia, and
$\mu_\mathrm{mod}(z_i)$ is the theoretical distance modulus
\begin{equation}
\mu_\mathrm{mod}(z_i)=5\log_{10}\frac{d_L(z_i)}{\mathrm{Mpc}}+25.
\end{equation}
To determine the likelihood of the parameters $B_s$ and $n$, we
marginalize the likelihood function $L=\exp(-\chi^2/2)$ over $h$
and $\Omega_b$. We adopt Gaussian priors such that $h=0.72\pm
0.08$ from the Hubble Space Telescope Key Project~\cite{WF} and
$\Omega_bh^2=0.0214\pm 0.0020$ from the observed abundances of
light elements together with primordial
nucleosynthesis~\cite{KTSOL}. The results of our analysis for the
VCG model are displayed in Fig.~\ref{fig:SN}. We show 68.3\%,
95.4\% and 99.7\% confidence level contours in the ($B_s$, $n$)
plane. The best-fit model parameters and marginalized $1\sigma$
error bars are $B_s=0.223^{+0.057}_{-0.059}$ and
$n=-3.0^{+2.4}_{-6.2}$ with $\chi^2_{\rm min}=174.253$. The
results show that the two parameter $B_s$ and $n$ are highly
degenerate. It is interesting to note that the dark energy
component with $w<-1$ is favored, which allows the possibility
that the dark energy density in increasing with time.

%%=======================section 3===========================
\section{Constraints from the X-ray Gas Mass Fraction of
Galaxy Clusters}

The matter content of rich clusters of galaxies is thought to
provide a fair sample of the matter content of the universe as a
whole. The observed ratio of the baryonic to total mass in
clusters should therefore closely match the ratio of the
cosmological parameter $\Omega_b/\Omega_m$, where $\Omega_b$ and
$\Omega_m$ are the mean baryon and total mass densities of the
universe in units of the critical density. The combination of
robust measurements of the baryonic mass fraction in clusters with
accurate determinations of $\Omega_b$ from cosmic nucleosynthesis
calculations can therefore be used to determine
$\Omega_m$~\cite{SDMW}. The measurements of the apparent redshift
dependence of the baryonic mass fraction can also, in principle,
be used to constrain the geometry and dark energy density of the
universe~\cite{SSUP}. The first successful application of such a
test was carried out by Allen {\it et al.} using a small sample of
X-ray luminous, dynamically relaxed clusters with precise mass
measurements, spanning the redshift range $0.1<z<0.5$~\cite{ASF}
(see also~\cite{zhu}). Recently Allen {\it et al.} present a
significant extension and obtained a tight constraint on the mean
matter density and dark energy equation of state parameter. The
clusters sample is significantly larger and includes 26 X-ray
luminous, dynamically relaxed systems spanning the redshift range
$0.07<z<0.9$~\cite{ASEFL}. We will use this database to constraint
the VCG model. To determine the confidence region of the model
parameters, we use the following $\chi^2$ function
\begin{equation}
\chi^2_{\mathrm{Xray}}=\sum_{i=1}^{26}\frac{\left[f_{\mathrm
gas}^{\mathrm
mod}(z_i;h,\Omega_b,B_s,n)-f_{\mathrm{gas},\,i}^{\mathrm
obs}\right]^2}{\sigma_{\mathrm{gas},\,i}^2}\,,
\end{equation}
where $f_{\mathrm{gas},\,i}^{\mathrm{obs}}$ is the measured X-ray
gas mass fraction $f_{\mathrm gas}$ with the defaut standard cold
dark matter (SCDM) cosmology, $\sigma_{\mathrm{gas},\,i}$ is the
symmetric root mean square errors, and
$f_{\mathrm{gas},\,i}^{\mathrm mod}$ is the model function
\begin{equation}
f_{\mathrm{gas}}^{\mathrm{mod}}(z_i)=\frac{b\Omega_b}
{(1+0.19h^{1/2})\Omega_{m}^{\mathrm{eff}}}
\left[\frac{h}{0.5}\frac{d_A^{\mathrm{SCDM}}(z_i)}{d_A^{\mathrm{
mod}}(z_i)}\right]^{3/2},
\end{equation}
where the bias factor $b=0.824\pm 0.089$~\cite{ASEFL} is a
parameter motivated by gas dynamical simulations, which suggest
that the baryon fraction in clusters is slightly depressed with
respect to the universe as a whole, and the effective matter
density parameter is
\begin{equation}
\Omega_{m}^{\mathrm{eff}}=\Omega_b+(1-\Omega_b)\sqrt{B_s}\,.
\end{equation}
Adopting Gaussian priors such that $h=0.72\pm 0.08$ and
$\Omega_bh^2=0.0214\pm 0.0020$, the 68.3\%, 95.4\% and 99.7\%
confidence level contours in the ($B_s$, $n$) plane are shown in
Fig.~\ref{fig:Xray}. The best-fit model parameters and
marginalized $1\sigma$ error bars are
$B_s=0.049^{+0.016}_{-0.015}$ and $n=0.5^{+1.0}_{-1.1}$ with
$\chi^2_{\rm min}=24.437$. The results favor the original
Chaplygin gas model with $n=0$.

%%=======================section 4===========================
\section{Combined Analysis, Conclusions and Discussions}

\begin{figure}
\begin{center}
\includegraphics[width=11cm]{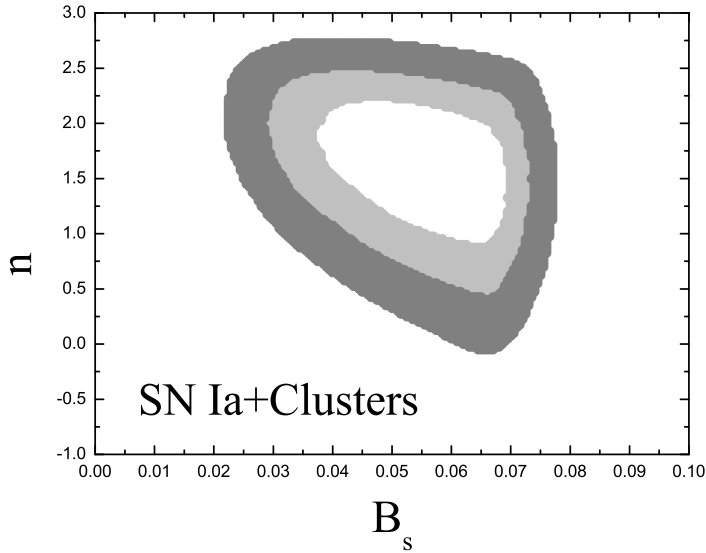}
\caption{Probability contours for $B_s$ versus $n$ are shown at
$1\sigma$, $2\sigma$, and $3\sigma$ when $\Omega_{\rm tot}=1$.
These constraints use the gold sample of 157 SN Ia data~\cite{AR}
and the X-ray gas mass fraction in 26 galaxy
clusters~\cite{ASEFL}.}
 \label{fig:SNXray}
\end{center}
\end{figure}

\begin{figure}
\begin{center}
\includegraphics[width=9.5cm]{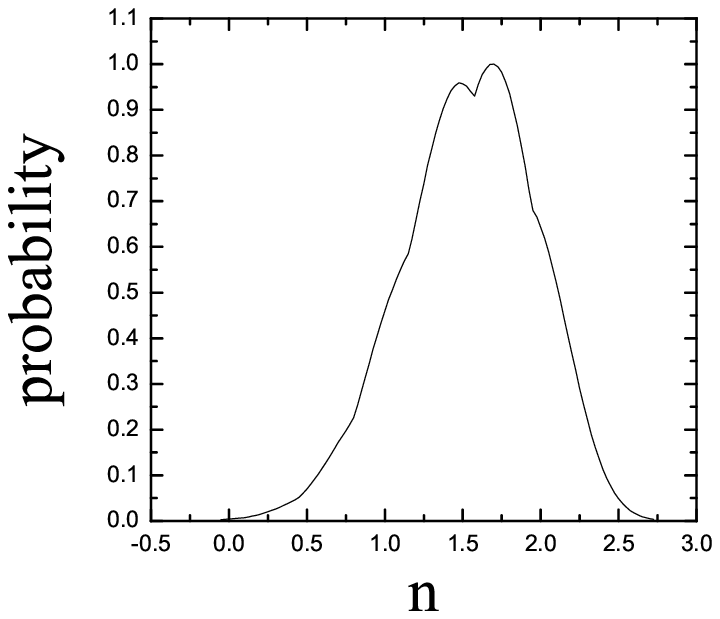}
\caption{Probability distribution for $n$ marginalized over $B_s$
obtained from the analysis of the combined SNIa$+f_{\rm gas}$ data
set in the VCG model.}
 \label{fig:pn}
\end{center}
\begin{center}
\includegraphics[width=9.5cm]{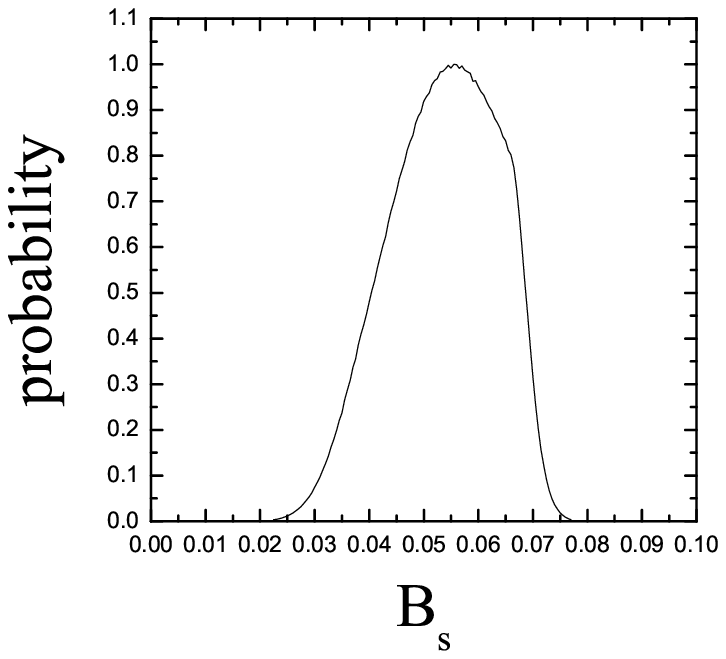}
\caption{Probability distribution for $B_s$ marginalized over $n$
obtained from the analysis of the combined SNIa$+f_{\rm gas}$ data
set in the VCG model.}
 \label{fig:pbs}
\end{center}
\end{figure}

Let us now consider a combined analysis of the constraints from
the gold sample of SN Ia data and the measurements of the X-ray
gas mass fraction in galaxy clusters. The fit is done by
minimizing the following $\chi^2$ function
\begin{equation}
\chi^2_{\mathrm{tot}}
=\chi^2_{\mathrm{SN}}+\chi^2_{\mathrm{Xray}}\,.
\end{equation}
Fig.~\ref{fig:SNXray} shows the constraints on $B_s$ and $n$
obtained from the analysis of the combined the SN Ia and the X-ray
gas mass fraction data set. We see that SNIa$+f_{\rm gas}$ data
set provides a remarkably tight constraint in the ($B_s$, $n$)
plane, with best fit values $B_s=0.055^{+0.013}_{-0.012}$
($1\sigma$ error bar) and $n=1.70^{+0.33}_{-0.52}$ ($1\sigma$
error bar) with $\chi^2_{\rm min}=205.457$. We find that the
original Chaplygin gas model is ruled out by the data at 99.7\%
confidence level. Fig.~\ref{fig:pn} and Fig.~\ref{fig:pbs} show
the probability distribution for $n$ and $B_s$, respectively,
marginalized over $B_s$ and $n$ in the spatially flat FRW
universe.

The VCG model, which plays the role of both dark energy and dark
matter in terms of a single component, deserves to explore its
various observational effects~\cite{SSKJD}. In this paper we have
focused our attention on the gold sample of SN Ia data and the
recent measurements of the X-ray gas mass fractions in galaxy
clusters. Adopting simple Gaussian priors of $h$ and
$\Omega_bh^2$, we have obtained stringent constraints on the two
parameters, $B_s$ and $n$, which describe the effective density
parameter of the dark matter component and the effective equation
of state of the dark energy component, respectively. The combined
analysis of these databases shows that the original Chaplygin gas
model is ruled out and the universe tends to be
quiessence-dominated rather than phantom-dominated at $3\sigma$
confidence level in the VCG scenario. It would be interesting to
investigate the evolution of density perturbations in this VCG
model.

\section*{Acknowledgements}
We are grateful to Zong-Hong Zhu and Adam G. Riess for helpful
discussions. We would like to thank Robert W. Schmidt for sending
us their compilation of the X-ray mass fraction data. This project
was in part supported by National Basic Research Program of China
under Grant No. 2003CB716300 and by NNSFC under Grant No.
90403032.

\end{document}